\documentclass[journal]{IEEEtran}
\usepackage{graphicx}
\usepackage{booktabs}
\usepackage{subfigure}
\usepackage{overpic}
\usepackage{amsmath}
\usepackage{flushend}
\usepackage{amssymb}
\usepackage{multirow}
\usepackage{makecell}
\usepackage{listings}
\usepackage{algorithm}
\usepackage{fontawesome}
\usepackage{color, soul}
\usepackage[absolute,overlay]{textpos}                    
\usepackage[table]{xcolor}
\usepackage{hyperref}
\hypersetup{
	colorlinks=true,
	linkcolor=blue,
	urlcolor=blue,
	citecolor=blue
}
\usepackage{colortbl}
\usepackage[dvipsnames]{xcolor}
\definecolor{bg}{HTML}{e0f1ff}

\begin{document}

\title{Wavelet-Assisted Mamba for Satellite-Derived Sea Surface Temperature Super-Resolution}
\author{
Wankun Chen, 
Feng Gao, \emph{Member}, \emph{IEEE},
Yanhai Gan, 
Jingchao Cao, \\
Junyu Dong, \emph{Member}, \emph{IEEE},
Qian Du, \emph{Fellow}, \emph{IEEE}
\thanks{This work was supported in part by the National Science and Technology Major Project under Grant 2022ZD0117201, in part by the Natural Science Foundation of Shandong Province under Grant ZR2024MF020, and in part by the Fundamental Research Funds for the Central Universities 202572015. \textit{(Corresponding author: Feng Gao and Junyu Dong)}

Wankun Chen and Junyu Dong are with the State Key Laboratory of Physical Oceanography, and also with the Frontiers Science Center for Deep Ocean Multispheres and Earth System, Ocean University of China, Qingdao 266100, China.

Feng Gao, Yanhai Gan, and Jingchao Cao are with the State Key Laboratory of Physical Oceanography, Faculty of Information Science and Engineering Ocean University of China, Qingdao 266100, China.

Qian Du is with the Department of Electrical and Computer Engineering, Mississippi State University, Starkville, MS 39762 USA.

Digital Object Identifier 10.1109/TGRS.2025.XXXXXXX}}

\markboth{IEEE TRANSACTIONS ON GEOSCIENCE AND REMOTE SENSING}{Shell}

\maketitle

\begin{abstract}
Sea surface temperature (SST) is an essential indicator of global climate change and one of the most intuitive factors reflecting ocean conditions. Obtaining high-resolution SST data remains challenging due to limitations in physical imaging, and super-resolution via deep neural networks is a promising solution. Recently, Mamba-based approaches leveraging State Space Models (SSM) have demonstrated significant potential for long-range dependency modeling with linear complexity. However, their application to SST data super-resolution remains largely unexplored. To this end, we propose the Wavelet-assisted Mamba Super-Resolution (WMSR) framework for satellite-derived SST data. The WMSR includes two key components: the Low-Frequency State Space Module (LFSSM) and High-Frequency Enhancement Module (HFEM). The LFSSM uses 2D-SSM to capture global information of the input data, and the robust global modeling capabilities of SSM are exploited to preserve the critical temperature information in the low-frequency component. The HFEM employs the pixel difference convolution to match and correct the high-frequency feature, achieving accurate and clear textures. Through comprehensive experiments on three SST datasets, our WMSR demonstrated superior performance over state-of-the-art methods. Our codes and datasets will be made publicly available at \url{https://github.com/oucailab/WMSR}.

\end{abstract}

\begin{IEEEkeywords}
Image Super-Resolution,
Sea Surface Temperature, 
State-Space Model, 
Image Restoration,
Discrete Wavelet Transform,
Pixel Difference Convolution.
\end{IEEEkeywords}

\IEEEpeerreviewmaketitle

\section{Introduction}

% 第一段：讲SST非常重要

\IEEEPARstart {O}{cean} is an integral component of the global climate system and plays a crucial role in the Earth's energy balance. Better understanding and monitoring the Earth's energy balance requires high-quality data. For instance, sea surface temperature (SST) from the satellite sensors or numerical modeling can reflect the overall warming trend in the global climate system \cite{Prochaska23tgrs}. Higher SST commonly leads to severe storms and weather events, including tropical cyclones, which draw energy from the ocean's surface to intensify. Furthermore, increased SSTs are associated with marine heatwaves, which have devastating effects on local ecosystems and are sometimes referred to as ``wildfires of the ocean" \cite{jbk22db}. Therefore, understanding and monitoring SSTs are crucial for assessing the impacts of climate change and predicting extreme weather events, as well as for comprehending the broader implications on global climate systems and marine life \cite{wentz00science} \cite{sby24grsl} \cite{xl24grsl}.

% 第二段：介绍 SST 可以从数值模式和遥感数据获取

Typically, SST data can be obtained through numerical modeling and satellite sensors \cite{4778913}. Numerical modeling is based on dynamics and state equations that incorporate various physical, chemical, and biological parameters and their intricate relationships \cite{COURTOIS201760}. The intricate physical dynamics of ocean models are characterized by partial differential equations, which are computationally expensive to solve. Consequently, despite the current computing capabilities, conducting long-term, high-resolution simulations to acquire detailed data remains infeasible \cite{physical}. Besides numerical modeling, satellite-derived SST data have been widely utilized to study the ocean circulation and atmosphere-ocean interactions \cite{meng23tgrs}. However, the trade-off between spatial resolution and revisit time makes the SST data products commonly only have a moderate resolution (25 km for microwave-based SST data and 1$\sim$4 km for infrared-based SST data) \cite{Prochaska23tgrs}. Therefore, there is ongoing interest in enhancing the resolution of SST data for long-term ocean monitoring in detail \cite{lloyd22tgrs}. To obtain high-resolution satellite-derived SST data, the super-resolution (SR) is a new and effective way to generate high-spatial resolution SST data from low-resolution data with the same coverage \cite{sr23review}. In this paper, we mainly focus on developing an accurate super-resolution method for satellite-derived SST data.

% 下面开始描述如何进行超分

In contrast to the ocean numerical model, deep learning-based super-resolution methods have the strong capacity of learning complex patterns and structures from SST data \cite{zjy24tip} \cite{xpc24tip}. These methods can generate more accurate and realistic high-resolution outputs. Furthermore, these methods have relatively lower computational costs than ocean numerical models \cite{zrt23rs}. Many deep learning-based super-resolution methods have been proposed and have demonstrated remarkable progress. Convolutional neural network (CNN) \cite{dong14eccv} \cite{yan22tmm}, and attention mechanism \cite{zhang18eccv} have been widely employed for natural image super-resolution. These CNN-based super-resolution methods commonly leverage an attention mechanism to emphasize informative features. However, the receptive fields of these methods are limited and can hardly capture long-range feature dependencies. 

Inspired by vision Transformer \cite{dos20vit}, many Transformer-based super-resolution methods have been proposed \cite{swinir} \cite{cyz23spl} \cite{lqg24tmm} \cite{hjf22grsl}. These methods increase the receptive fields by leveraging the global feature interactions via the self-attention mechanism.  Nevertheless, training Transformer-based super-resolution methods for high-resolution images presents a significant challenge due to their quadratic complexity in relation to token size \cite{hdc23iccv}. Modeling long-range feature dependencies more efficiently is garnering more attention from researchers.

% 提出来当前有一种新的 MAMBA 方法，可以有效解决当前问题

Recently, structured State-Space Model (SSM) has emerged as an effective tool for modeling long sequences with linear complexity \cite{js23iclr} \cite{lyp24tgrs} \cite{mxp24grsl} \cite{zsj24tgrs}. Particularly, the improved SSM with selective scanning mechanism and efficient hardware design, Mamba \cite{gu23mamba} has demonstrated excellent performance for long-term dependency modeling. Compared with traditional Transformer-based methods, Mamba reformulates the attention mechanism so that it scales linearly with the sequence length, the computational costs are significantly reduced \cite{mambair2024eccv}. The selective scan mechanism enabling enhanced processing of complex spatiotemporal sequences such as ocean remote sensing data \cite{gao2025tgrs}. Its efficacy is evidenced by recent applications in geospatial data reconstruction \cite{ma2024grsl} \cite{chen2024grsl} \cite{chen2024tgrs}. Crucially, this approach allows any pixel to aggregate contextual information from multiple directions \cite{chen2024tgrs}, facilitating comprehensive feature extraction. However, Mamba has rarely explored to solve the SST data super-resolution problem, which motivates us to explore the potential of the selective scan mechanism to enhance the SST data super-resolution performance.

% 指出来，设计基于 Mamba 的SST超分方法存在两个挑战

It is a non-trivial task to develop robust SST data super-resolution method based on SSM, due to the following two challenges: \textit{\textbf{1) Frequency global degradation perception.}} Mamba implements selective scanning along one-dimensional spatial sequences, thus limiting its ability to capture frequency-domain characteristics. Incorporating such features could enhance SST super-resolution performance. Therefore, how to adaptively integrate the frequency information in Mamba poses the main challenge for us. \textit{\textbf{2) Spatial details need to be enhanced.}} Most existing methods use Feed-Forward neural Network (FFN) to enhance the non-linear feature transformation. Channel attention is commonly used to explore inter-channel relationships, and it lacks explicit consideration of spatial details. Consequently, how to incorporate the spatial details in the FFN is a tough challenge.

% 指出来，为了解决上述两个问题，本文是如何做的

To mitigate the above problems, we propose a \textbf{W}avelet-assisted \textbf{M}amba \textbf{S}uper-\textbf{R}esolution (\textbf{WMSR}) framework for satellite-derived SST data, which combines the frequency feature modeling and SSM. Specifically, to adaptively integrate the frequency information into Mamba, we design \textit{Wavelet-Assisted Mamba (WAM)} block. In each WAM block, the input features are divided into high- and low-frequency features using the discrete wavelet transform. For the low-frequency features, we propose \textit{Low-Frequency State Space Module (LFSSM)}, which introduces 2D-SSM to capture global characteristics of the input data. To enhance the spatial details, we design \textit{High-Frequency Enhancement Module (HFEM)}, which uses the pixel difference convolution to enhance the high-frequency features. Pixel differential convolution calculates the pixel differences in the image and then inputs them into the convolution kernel for convolution to generate the output. This pixel pair difference calculation strategy can explicitly encode high-frequency prior information into the model, further learn beneficial gradient information, and  improves the reconstruction ability of spatial details.

The contributions of our WMSR are summarized as follows:

\begin{itemize}

\item We propose a conceptually novel yet simple framework termed WMSR for satellite-derived SST data super-resolution, which combines the frequency feature modeling and SSM. To the best of our knowledge, this is the first work that attempts to solve the SST data super-resolution problem by leveraging the selective scan mechanism of SSM.

\item We present LFSSM to leverage 2D-SSM to capture global information of the input data. The global modeling capabilities of SSM are exploited to preserve the critical temperature information in the low-frequency component. In addition, we design HFEM to enhance the spatial details via pixel difference convolution, which can highlight the subtle changes and variations, thereby reconstructing precise textural details through targeted feature rectification.

\item  We conduct comprehensive experiments on three SST datasets. The experimental results show that the proposed WMSR framework achieves superior super-resolution performance over the state-of-the-art methods. In addition, we will release our codes and datasets to facilitate other researchers. 

\end{itemize}

\section{Related Works}
\subsection{Natural Image Super-Resolution}

There are three main categories of natural image super-resolution approaches: mathematical interpolation-based methods \cite{chazhi}, image reconstruction-based methods \cite{chong}, and learning-based methods \cite{dl}. Among them, learning-based methods are the most successful super-resolution approaches due to their remarkable spatial feature extraction capability \cite{dlm}. Dong et al. \cite{srcnn} put forward the Super-Resolution Convolutional Neural Network (SRCNN). This represents a pioneering endeavor that makes use of a three-layer CNN framework to map low-resolution images to their high-resolution equivalents. Kim et al. \cite{vdsr} proposed Very Deep Convolutional Networks (VDSR), which introduces residual learning into super-resolution to achieve better accuracy and visual improvements. Later, Shi \cite{shi16cvpr} proposed a more effective sub-pixel convolution layer instead of transposed convolution and achieved better performance. Dai et al. \cite{dai19cvpr} presented a second-order attention network to capture long-distance spatial contextual information. To further improve the computational efficiency, Luo et al. \cite{luo20eccv} proposed LatticeNet, which incorporated the faster Fourier transform to construct a lattice. Such CNN-based approaches demonstrate significant advancements in image super-resolution tasks.

Recently, Transformer-based approaches have replaced CNN-based methods to elevate super-resolution performance. SwinIR \cite{swinir} uses the Swin Transformer with window-based attention for image restoration. Meanwhile, permuted self-attention is employed in SRFormer \cite{zhou23iccv} for image super-resolution. Permuted self-attention strikes an appropriate balance between the channel-wise and spatial-wise attention, and achieves competitive results. Yoo et al. \cite{yoo23wacv} proposed a super-resolution network, which aggregates local features from CNNs and long-range multi-scale dependencies from the Transformer. ELAN \cite{elan} employs self-attention computed in different window sizes to collect the correlations among long-range pixels. 

Researchers have successfully applied these methods to super-resolution of remote sensing images as well, and have achieved notable results. MSWAGAN \cite{wang2024tgrs} combines multi-scale sliding window attention with the standard Transformer, taking into account both local complex features and global long-range dependencies. ConvFormerSR \cite{li2024tgrs} effectively enhances the quality of super-resolution reconstruction of multispectral remote sensing images by fusing CNN and Transformer. LGC-GDAN \cite{liht2024tgrs} effectively captures the global and local features of remote sensing images through a dual-branch structure of context and edge. DBSAGAN \cite{song2024lgrs} has achieved good results by organically combining the dual-branch attention mechanism with frequency domain constraints.
Xiao et al. \cite{xy25tmm} integrates the SSM for remote sensing image super-resolution. It uses a multi-level fusion framework equipped with the frequency selection module and vision SSM for effective spatial-frequency fusion. TTST \cite{xy24tip} is a lightweight Transformer-based super-resolution method. It introduces a residual token selective mechanism to dynamically filter out irrelevant tokens. It achieves satisfying super-resolution performance while using less computational cost and parameters. In addition, some methods have achieved good performance by changing the network architecture, such as U-shaped network structure \cite{jiang2022grsl}, multi-scale feature \cite{wang2023grsl}, and domain matching \cite{min2024grsl}.

Although these methods achieve competent super-resolution performance, they exhibit limited efficacy for sea surface temperature (SST) data reconstruction. Unlike natural images with rich textural details, SST datasets demonstrate lower textural complexity and stronger spatial continuity.  Consequently, our framework prioritizes  on modeling the critical temperature information in the low-frequency component via SSM.

\begin{figure*}[]
  \centering
  \includegraphics[width=6in]{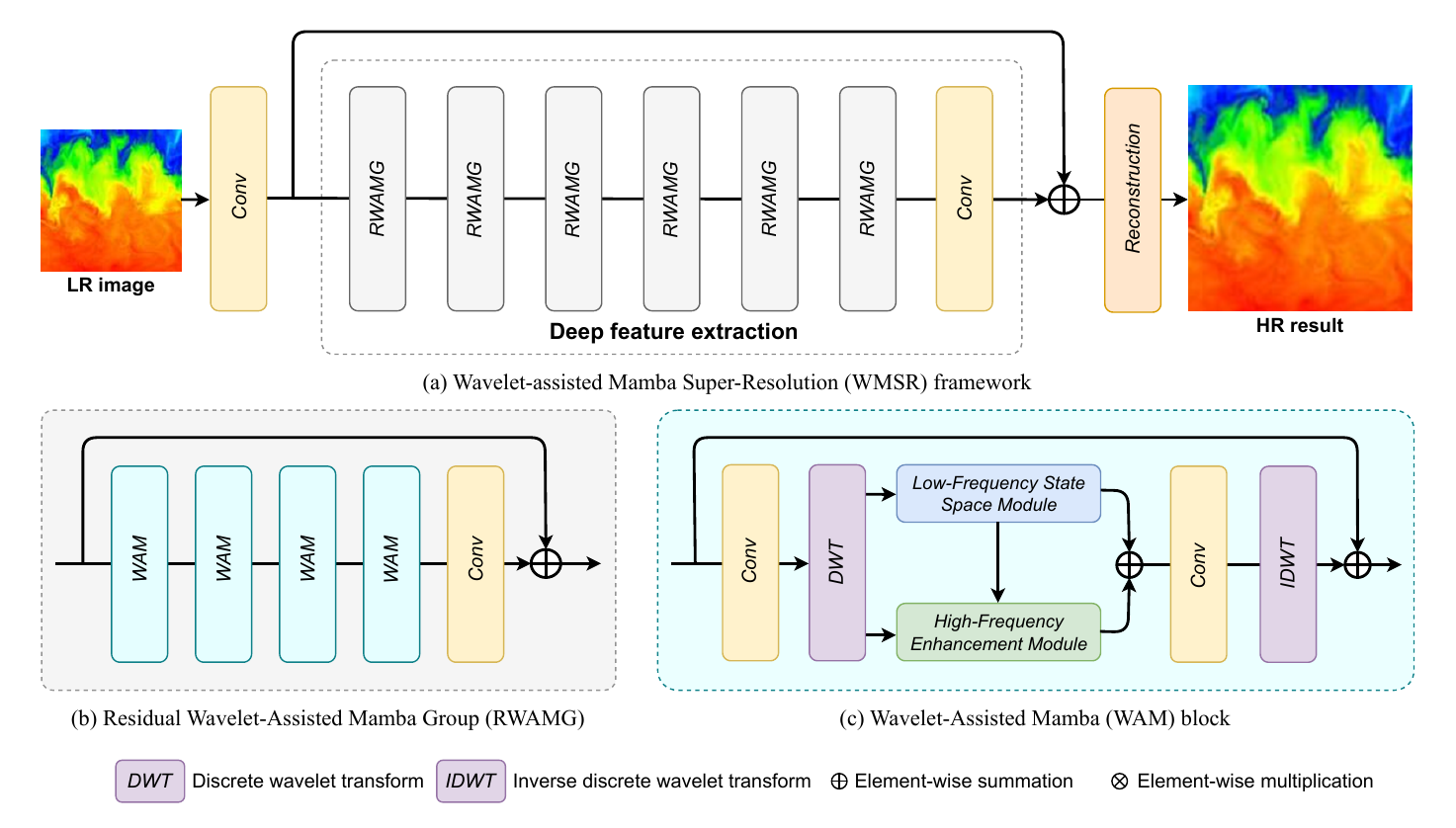}
  \caption{Overview of our proposed Wavelet-assisted Mamba Super-Resolution (WMSR) framework for SST data super-resolution. (a) illustrates the overall architecture of WMSR. The WMSR includes several stacked Residual Wavelet-Assisted Mamba Groups (RWAMG). (b) displays the structure of the RWAMG. It contains several Wavelet-Assisted Mamba (WAM) blocks. (c) shows the details of the WAM block. Discrete wavelet transform is employed to separate the low- and high-frequency components. They are fed into the Low-Frequency State Space Module (LFSSM) and High-Frequency Enhancement Module (HFEM) for feature extraction, respectively.} 
  \label{fig_network}
\end{figure*}

\subsection{Sea Surface Temperature Super-Resolution Based on Deep Learning}

There has recently been growing interest in enhancing the resolution of SST via deep-learning techniques. Ducournau et al. \cite{ducournau16prrs} designed SRCNN network for SST data super-resolution. Ping et al. \cite{ping21jstars} proposed an oceanic data reconstruction network, which employed multi-scale feature extraction and multi-receptive field mapping for SST reconstruction. Izumi et al. \cite{rs13183568} utilized the enhanced super-resolution generative adversarial network (ESRGAN) to perform super-resolution on SST data. Additionally, they compared various classical methods and analyzed the distinct effects of GAN and CNN on the SST data super-resolution tasks. Kim et al. \cite{kim23ija} proposed a GAN-based spatio-temporal learning method for SST data super-resolution. Zou et al. \cite{zou23rs} presented a transformer-based SST reconstruction model, which incorporates the transformer block and the residual block for cross-scale feature learning.

Our method differs from preceding endeavors through two innovations. Firstly, we employ the selective scan mechanism of SSM to solve the SST data super-resolution task. Secondly, we use the pixel difference convolution to enhance the high-frequency features.

\subsection{State Space Models}

State Space Models (SSM) \cite{gu21ssm} have gained significant attention recently. They can model long-range dependencies while exhibiting linear scalability with sequence length. Mehta et al. \cite{mehta22long} incorporated gating units into SSM and enhanced the performance in long-range language modeling. Gu et al. \cite{gu23mamba} proposed Mamba, which is a data-dependent SSM featuring a selective mechanism. It has surpassed Transformer in natural language modeling while maintaining linear scalability with input length. Recently, some pioneering works have used Mamba for various remote sensing image understanding tasks, including change captioning \cite{liu24rsca}, image segmentation \cite{liu24cmunet}, image dehazing \cite{zhou24rsde}, etc. These methods have achieved promising performance. However, Mamba remains unexplored for SST data super-resolution, which motivates us to investigate vision Mamba to enhance the performance of SST data super-resolution.

\section{Methodology}

In this section, we first provide preliminaries of the Mamba. Then we describe the implementation details of our WMSR, followed by a detailed description of the proposed Wavelet-Assisted Mamba (WAM) block, the Low-Frequency State Space Module (LFSSM), and High-Frequency Enhancement Module (HFEM). Finally, we provide descriptions of the loss function. 

\subsection{Preliminaries}
\textbf{State Space Model} (SSM) is a linear time-invariant system that maps an input sequence $x(t)\in\mathbb{R}^N$ to an output sequence $y(t)\in\mathbb{R}^N$. The system can be formulated as a linear Ordinary Differential Equation (ODE) as follows:
\begin{equation}
    h'(t)= \mathbf{A}h(t)+ \mathbf{B}x(t),
\end{equation}
\begin{equation}
    y(t)=\mathbf{C}h(t)+\mathbf{D}x(t),
\end{equation}
where $h'(t)\in \mathbb{R}^N$ is the implicit latent state, $N$ is the number of states. $\mathbf{A}\in \mathbb{R}^{N\times N}$ is the state transition matrix. $\mathbf{B}$ and $\mathbf{C}$ are projection matrices. $\mathbf{D}$ is the residual connected operation. 

In real applications, these continuous equations need to be discretized for computational tractability and alignment with the input data. A timescale parameter $\Delta$ is incorporated to convert the continuous parameters $\mathbf{A}$, $\mathbf{B}$ into discrete parameters $\overline{\mathbf{A}}$, $\overline{\mathbf{B}}$ as follows:
\begin{equation}
  \overline{\mathbf{A}} = \mathrm{exp}({\Delta \mathbf{A}}),
  \label{eq3}
\end{equation}
\begin{equation}
  \overline{\mathbf{B}} = (\Delta \mathbf{A})^{-1}(\mathrm{exp}(\Delta\mathbf{A})- \mathbf{I}) \cdot \Delta \mathbf{B}.
  \label{eq4}
\end{equation}

After discretization, the ODE of SSM can be computed as follows: 
\begin{equation}
   h_t = \overline{\mathbf{A}}h_{t-1} + \overline{\mathbf{B}}x_{t}, 
   \label{eq5}
\end{equation}

\begin{equation}
   y_t = \mathbf{C}h_{t} + \mathbf{D}x_t
   \label{eq6}
\end{equation}

\textbf{Selective Scan Mechanism.} Traditional SSM employ linear time-invariant frameworks, which means that the projection matrices remain fixed and unaffected by variations in the input sequence, those inherent rigidity fundamentally hinder local dependency modeling within sequential data structures. To alleviate this limitation, Mamba \cite{gu23mamba} proposes a solution where the parameter matrices become input-dependent. In this way, SSM can better manage complex sequences, potentially enhancing their capability through the transformation into linear time-varying systems.

{The 2D Selective Scan Module (2D-SSM) decomposes the spatial reasoning into multiple directional scans to comprehensively capture contextual information, the input feature map \(\mathbf{X} \in \mathbb{R}^{H \times W \times C}\) is unfolded into four separate 1D sequences along four distinct directions: top-left to bottom-right, top-right to bottom-left, and their reverse orders, each providing a unique spatial perspective. Each sequence is processed independently by a 1D selective state space model (SSM). This is implemented via a linear projection of the input sequence. The discretization of the continuous SSM follows Eq. (\ref{eq3}-\ref{eq6}), enabling efficient recursive computation. The output sequences from the 2D-SSM are folded back to their original 2D shapes to produce the final output \(\mathbf{F}_{out}\), which integrates information from all spatial pathways.}

\subsection{Overall Framework of the WMSR}

As depicted in Fig. \ref{fig_network}, the proposed WMSR consists of three parts: local feature extraction, deep feature extraction, and high-quality reconstruction. In local feature extraction, given a low resolution image $I_{LR} \in \mathbb{R}^{H \times W}$, a feature representation $F_S \in \mathbb{R}^{H\times W\times C}$ is produced via a convolution layer, where $H\times W$ is the spatial resolution, and $C$ is the number of channels. Next, $F_S$ are fed into the deep feature extraction stage to acquire the deep feature $F_D$. This stage consists of several stacked Residual Wavelet-Assisted Mamba Groups (RWAMG).

Each RWAMG contains several WAM blocks. In each WAM block, the wavelet transform is used to assist Mamba to exploit both high-frequency and low-frequency information. Moreover, an additional convolution layer is employed at the end of each group to refine features extracted from WAM. Then, the shallow feature $F_S$ and deep feature $F_D$ are combined via element-wise summation. Finally, the obtained features are further refined by pixel shuffle operation to generate the final HR estimation $I_{SR}$. 

As illustrated in Fig. \ref{fig_network}, in the WAM module, {we apply the Discrete Wavelet Transform (DWT) using the Haar wavelet to extract low-frequency and high-frequency components. SST fields are predominantly composed of extensive regions with smooth thermal gradients (low-frequency components), interspersed with localized, sharp discontinuities at oceanic fronts and eddy boundaries (high-frequency components). The Haar wavelet, with its compact support and rectangular basis functions, excels at representing such piecewise-constant signals and abrupt transitions.}

The kernel functions for the Haar wavelet are defined as $f_l =  \tfrac{1}{\sqrt{2}}[1,\ 1],f_h = \tfrac{1}{\sqrt{2}}[1,\ -1]$. The information of each frequency is extracted as follows: 

\begin{equation}
\begin{split}
\mathrm{LL} &= f_l \ast (f_l \ast I)^T \\
\mathrm{LH} &= f_h \ast (f_l \ast I)^T \\ 
\mathrm{HL} &= f_l \ast (f_h \ast I)^T \\
\mathrm{HH} &= f_h \ast (f_h \ast I)^T
\end{split}
\end{equation}
where $\mathrm{LL}$ is the low-frequency component of the input data $I$, while $\{\mathrm{LH}, \mathrm{HL}, \mathrm{HH}\}$ are the high-frequency components.

The low-frequency component $\mathrm{LL}$ is fed into the LFSSM for enhanced low-frequency feature extraction. The high-frequency component $\{\mathrm{LH}, \mathrm{HL}, \mathrm{HH}\}$ is fed into the HFEM for enhanced high-frequency feature extraction. Features from LFSSM and HFEM are fused via element-wise multiplication. Then, a convolution layer is employed for feature refinement, and Inverted Discrete Wavelet Transform (IDWT) is employed for reconstruction. LFSSM and HFEM are the critical components in the proposed WAM, and will be detailed in the following subsections.

\subsection{Low-Frequency State Space Module (LFSSM)}

The Low-Frequency State Space Module (LFSSM) is employed to extract and model low-frequency information from the spatial domain. Given the input low-frequency feature $\mathbf{F}_l \in \mathbb{R}^{H\times W \times C}$, we initially use Layer Normalization (LN), followed by the 2D Vision State Space Module (VSSM) to capture the long-term feature dependencies as follows:

\begin{equation}
    \mathbf{Z}=\text{VSSM}(\text{LN}(\mathbf{F}_l)) + \mathbf{F}_l,
\end{equation}
\begin{equation}
    \mathbf{F}_{o}=\text{GatedFFN}(\mathbf{Z})+\mathbf{Z},
\end{equation}
where VSSM($\cdot$) denotes the VSSM function, and LN($\cdot$) denotes the operation of layer normalization. Gated FNN($\cdot$) denotes the Gated Feed-Forward Network (FFN) for non-linear feature transformation.

We employ the Gated FFN for nonlinear feature transformation to regulate the information flow, which enables individual channels to concentrate on fine details that complement those from other layers. Specifically, the input feature $\mathbf{Z} \in \mathbb{R}^{H\times W \times C}$ is handled by layer normalization and depth-wise convolution. Then, the obtained feature are split along the channel dimension into two parts $\mathbf{Z}_1$, $\mathbf{Z}_2 \in \mathbb{R}^{H\times W \times \frac{C}{2}}$. The output is then calculated by non-linear gating as $\mathbf{Z}_o = \sigma(\mathbf{Z}_1)\odot \mathbf{Z}_2$, where $\sigma(\cdot)$ denotes the sigmoid activation. $\mathbf{Z}_o$ is the output of the Gated FNN.

\begin{figure}
    \centering
    \includegraphics[width=3in]{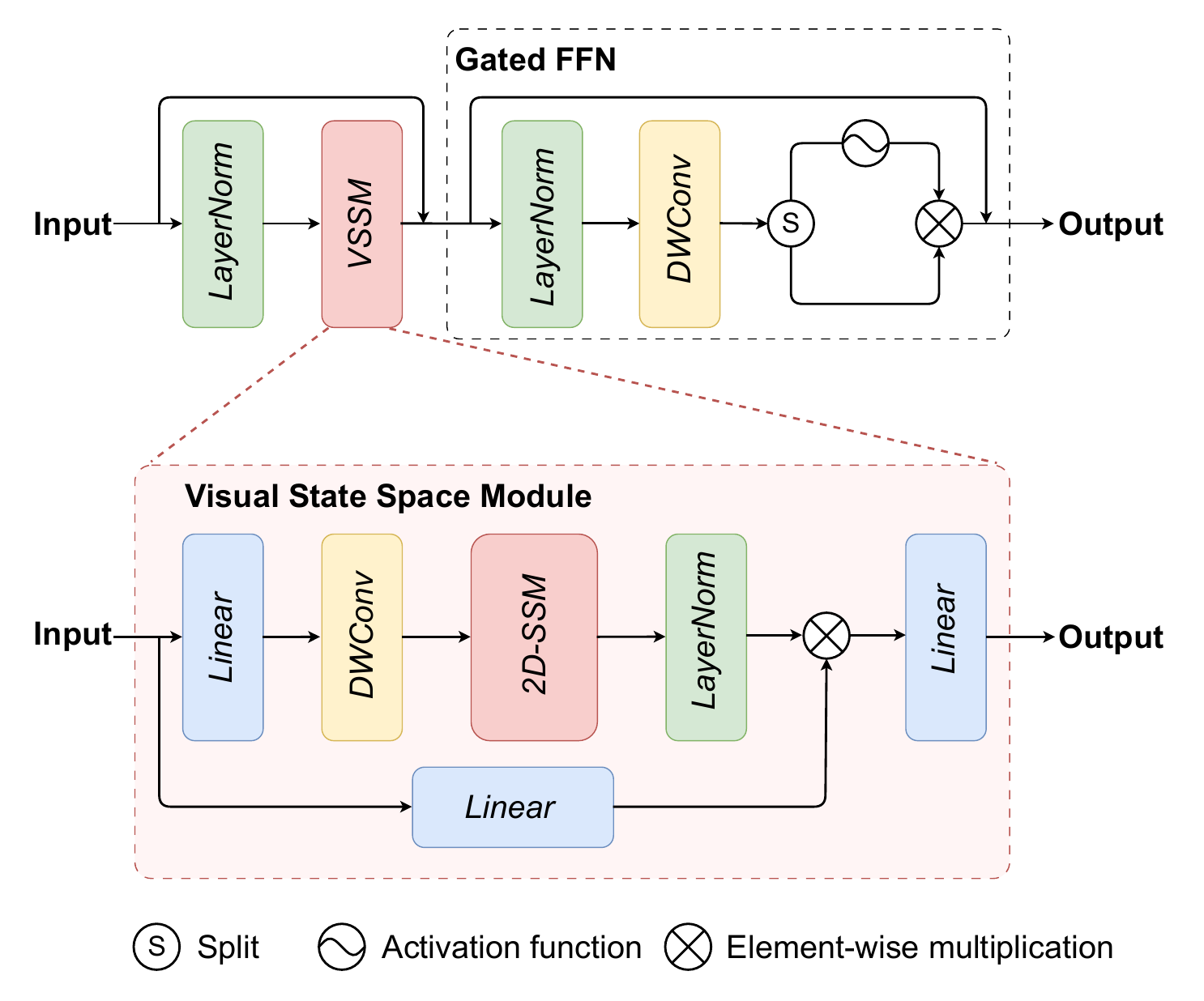}
    \caption{Details of the Low-Frequency State Space Module (LFSSM). The Vision State Space Module (VSSM) is employed to capture the long-term feature dependencies. Gated FFN is used for nonlinear feature transformation to regulate the information flow.}
    \label{fig_lfssm}
\end{figure}

We incorporate the VSSM into low-frequency feature extraction. Specifically, the input feature $\mathbf{X}$ is handled by two parallel branches. In the first branch, a linear layer expands the feature channel. After that, depth-wise convolution and SiLU activation are employed for feature extraction. 2D-SSM and layer normalization are also employed for feature modeling. In the second branch, a linear layer and SiLU activation are used. Finally, element-wise multiplication combines the outputs of both branches. Channels are reduced back, and $\hat{\mathbf{X}}$ is generated with the same input channel dimension. The computation process is formulated as follows:

\begin{equation}
    \mathbf{X}_1=\textrm{LN}(\textrm{2D-SSM}(\textrm{SiLU}(\textrm{DWConv}(\textrm{FC}(\mathbf{X}))))),
\end{equation}
\begin{equation}
    \mathbf{X}_2=\textrm{SiLU}(\textrm{FC}(\mathbf{X})),
\end{equation}
\begin{equation}
    \hat{\mathbf{X}}=\textrm{FC}(\mathbf{X}_1 \odot \mathbf{X}_2),
\end{equation}
where DWConv($\cdot$) denotes the depth-wise convolution, and FC($\cdot$) denotes the linear projection. $\odot$ denotes the element-wise multiplication.

\subsection{High-Frequency Enhancement Module (HFEM)}

\begin{figure}
    \centering
    \includegraphics[width=2.7in]{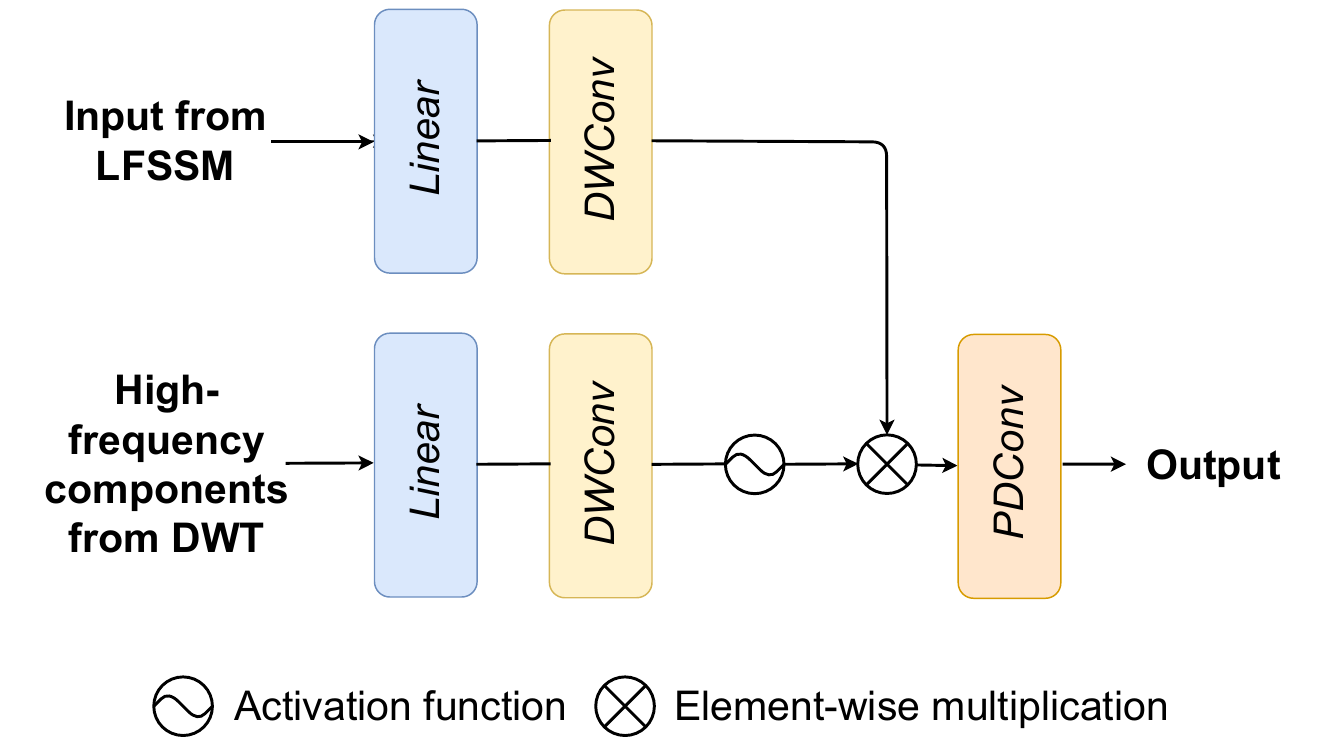}
    \caption{Details of the High-Frequency Enhancement Module (HFEM). Low-frequency features are integrated into high-frequency features via gated fusion. Pixel difference convolution is used for high-frequency feature enhancement.}
    \label{fig_hfem}
\end{figure}

In this section, to accomplish the extraction of high-frequency features, we employ the differential convolution to amplify the gradient information, while using the re-parameterization method to minimize the parameters and computational burden. The structure of the HFEM module is shown in Fig. \ref{fig_hfem}.

The features from LFSSM $\mathbf{F}_{l}$ and high-frequency features $\mathbf{F}_{wh}$ from DWT are fused via gated fusion as follows:
\begin{equation}
    \mathbf{F}_g= Gate(
    \textrm{DWConv}(\textrm{Linear}(\mathbf{F}_{l})), \mathbf{F}_{wh}),
\end{equation}
where Linear($\cdot$) denotes the fully connected layer, DWConv($\cdot$) denotes the $3\times3$ depth-wise convolution. The gate unit $Gate(\cdot)$ is formulated as:
\begin{equation}
    Gate(\mathbf{X}, \mathbf{Y})=
    \sigma (\textrm{PDConv}(\textrm{Linear}(
    \mathbf{Y})) ) \odot \mathbf{X}),
\end{equation}
where $\sigma(\cdot)$ refers to the sigmoid activation function, $\odot$ denotes the element-wise multiplication, and PDConv($\cdot$) denotes the pixel difference convolution.

%After that, $\mathbf{F}_g$ and low-frequency features from DWT $\mathbf{F}_{wl}$ are fused via gated fusion as:

%\begin{equation}
%\mathbf{F}_{out}=Gate(\textrm{DWConv}(\textrm{Linear}(\mathbf{F}_g, \mathbf{F}_{wl}))).
%\end{equation}

The pixel difference convolution proves to be effective for face anti-spoofing \cite{yzt21pami} and edge detection tasks \cite{sz21iccv}.{ Pixel difference convolution operates by enhancing pixel variations to elucidate unique properties, this strategy can explicitly encode high-frequency prior information, further learn beneficial gradient information. In this paper, four kinds of pixel difference convolution are employed, including Central Difference Convolution (CDC), Angular Difference Convolution (ADC), Horizontal Difference Convolution (HDC) and the Vertical Difference Convolution (VDC). To the best of our knowledge, it is the first time that we introduce pixel difference convolution to solve the SST super-resolution task.}

To illustrate, the CDC can be taken as an example: initially, a patch commensurate with the convolution kernel, a $3\times3$ matrix, is selected from the feature. Its central gradient is computed, and a differential operation is then applied.  Subsequently, upon having derived features from the aforementioned difference, a $3\times3$ convolution kernel is applied. This process can be described as:

\begin{equation}\begin{aligned}
y(p_{0})=\sum_{p_{n}\in \mathcal{R}}{w(p_{n})}\cdot (x(p_{0}+p_{n})-x(p_{0})),
\end{aligned}
\end{equation}
where $p_0$ denotes the current location in both input and output feature maps, and $p_n$ denotes the location in the local receptive field $\mathcal{R}$, and the local receptive field region  $\mathcal{R}$ is set to $3\times3$. %We create 8 pairs in the angular direction in the local patch, then the pixel differences obtained from the pairs can be convolved with the kernel by doing an element-wise multiplication with the kernel weights, followed by a summation, to generate the value in the output feature map. 

{For ADC, HDC and VDC, details can be found in \cite{sz21iccv}. They are implemented by re-arranging learned kernel weights to save computational burden and memory consumption.  Specifically,  each branch has its own trainable kernel parameters \( K_i \), where \( i \in \{1, 2, 3, 4, 5\} \). The outputs of all branches are summed together, and due to the linearity of the convolution operation, this summation is equivalent to a single convolution with a fused kernel \( k_f \):
\begin{equation}
\begin{aligned}
\text{PDConv}(\mathbf{X}) =\sum_{i=1}^{5}\mathbf{X} * k_{i} = \mathbf{X} * \left( \sum_{i=1}^{5} k_{i} \right) = \mathbf{X} * k_{f},
\end{aligned}
\end{equation}
where $k_i$, $i=1,\ldots, 5$ denote the kernels of vanilla convolution, CDC, ADC, HDC and VDC, respectively. $k_f$ is the fused kernel, which combines the parallel convolutions. }

{This fusion process is performed once after training, converting the multi-branch training-time architecture into a single, highly efficient convolution layer for deployment. This approach significantly reduces computational overhead while preserving the rich feature representation capabilities learned during training.}

\subsection{Loss Function}

During the training process, for given high-resolution SST data $I_{HR}$, where $N$ denotes the number of images, we use a $L_1$ reconstruction loss function as follows:
\begin{equation}
    \mathcal{L}_{rec}=||I_{SR}-I_{HR}||_1.
\end{equation}

In this work, we leverage frequency-domain information within SST data. Minimizing frequency discrepancies between original and restored images enhances super-resolution performance\cite{fpl}. Specifically, the Discrete Fourier Transform (DFT) is employed to transform the SST data from the spatial domain to the frequency domain as follows:

\begin{equation}
  F(u,v) = \cfrac 1 {HW} \sum_{x=0}^{H-1} \sum_{y=0}^{W-1} f(x,y) \cdot e^{-\cdot 2 \pi i \left( \frac{ux}{H} + \frac{vy}{W} \right)},
\end{equation}
where $H$ and $W$ denote the width and height of input data, with $(x, y)$ symbolizing the image pixel coordinates in the spatial domain. $f$ represents the corresponding pixel value at $ (x, y)$, while $(u, v)$ signifies the spatial frequency coordinates on the spectrum. $F$ is the complex frequency value, and $i$ is the imaginary unit. 

Let the high-resolution SST data in the frequency domain be $F_{HR}$, and the generated SST data be $F_{SR}$, we compute the frequency loss as follows:

\begin{equation}
 \mathcal{L}_{freq} = \omega(u,v)|F_{HR}(u,v)-F_{SR}(u,v) |^2,
\end{equation}
where $\omega=|F_{HR}(u,v)-F_{SR}(u,v)|$. Here we use a spectrum weight matrix that automatically ascertains the appropriate coefficients.

We employ a composite loss function $\mathcal{L}_{total}$, which combines reconstruction loss $\mathcal{L}_{rec}$ and frequency loss $\mathcal{L}_{freq}$ as follows:

\begin{equation}
    \mathcal{L}_{total}=\lambda_{rec} \mathcal{L}_{rec}+\lambda_{freq}\mathcal{L}_{freq}.
\end{equation}

By introducing frequency loss into SST data super-resolution, it can help the model to learn more information and achieve better reconstruction performance.

\section{Experimental Results and Analysis}

In this section, comprehensive SST data super-resolution experiments are performed on three datasets to evaluate the effectiveness of the proposed WMSR. We compare the proposed method with six state-of-the-art methods. These methods include Enhanced Deep Residual Networks (EDSR) \cite{edsr2017cvpr}, Context Reasoning Attention Network (CRAN) \cite{cran2021iccv}, and Dual Aggregation Transformer (DAT) \cite{dat2023iccv}, Spatially Adaptive Feature Modulation (SAFM) \cite{safmn2023iccv}, and State-Space Model-based method (MambaIR) \cite{mambair2024eccv}. Finally, ablation studies are conducted to verify the effectiveness of different components of the proposed network.

\subsection{Dataset Settings}

We conducted extensive experiments utilizing diverse sea surface temperature (SST) data sources, including the HYbrid Coordinate Ocean Model (HYCOM), the Optimum Interpolation Sea Surface Temperature (OISST) product, and the Group for High-Resolution Sea Surface Temperature (GHRSST) datasets. Among these, OISST and GHRSST represent remote sensing data, serving as direct observational sources of SST. They provide globally comprehensive spatial snapshots of SST derived from actual physical measurements. Conversely, HYCOM constitutes ocean model data, representing a physical-dynamical simulation of the ocean state that delivers temporally continuous estimates of oceanic conditions. 1,000 HR images were acquired per dataset, and downsampling using Bicubic. All data were partitioned into training and test subsets with a 4:1 ratio.

\textbf{HYCOM dataset} \cite{hycom}: HYCOM data were collected from January to December 2016, with a spatial resolution of 1/12° and a time step of 3 hours. The research area mainly of unfrozen oceans, namely the North Pacific (NP, 5°N--45°N, 140°E--180°E), the Atlantic Gulf of Mexico (AGM, 5°S--45°S, 35°W--75°W), the North Indian Ocean (NIO, 5°N--35°S, 50°W--90°W), and the Equatorial Warm Pool (EWP, 20°N--20°S, 120°W--160°W).

\textbf{OISST dataset} \cite{oisst}: The observation data OISST used in this study were downloaded from the National Oceanic and Atmospheric Administration website. The OISST data has a resolution of 1/4°, covering the period from January to December 2016. 

\textbf{GHRSST dataset} \cite{ghrsst}: Group for High Resolution Sea Surface Temperature (GHRSST) was established to foster an international focus and coordination for the development of a new generation of global, multi-sensor, high-resolution near real-time SST products. The data was collected on August 12, 2016, with a resolution of 0.01°.

\subsection{Training Details and Evaluation Metrics}

This experiment was implemented on an NVIDIA RTX 4080 GPU workstation using PyTorch 2.0. All convolutional layer weights were initialized with Kaiming normal distribution\cite{Kaiming}, while BatchNorm parameters were set to $\gamma=1$ and $\beta=0$. Fully-connected layers employed Xavier uniform initialization\cite{Xavier} - an effective strategy to mitigate vanishing gradient issues in deep networks. Training utilized the Adam optimizer ($\beta_1 = 0.9$, $\beta_2 = 0.999$, $\epsilon = 10^{-8}$) for 100 epochs with fixed $96 \times 96$ input patch size. The initial learning rate was set to $10^{-3}$, decaying by 50$\%$ every 20 epochs following a cosine annealing schedule.We adopted two evaluation metrics: Peak signal-to-noise ratio (PSNR) and Structural Similarity Index Measurement (SSIM, range: $[0, 1]$) to assess the effectiveness of the proposed WMSR in reconstructing high-resolution SST data.

\subsection{Hyperparameter Sensitivity Analysis}

This study employs a baseline architecture comprising four Residual Wavelet-Assisted Mamba Groups (RWAMGs), each containing four Wavelet-Assisted Mamba (WAM) modules. 

\begin{figure} [!h]
  \centering
  \includegraphics[width=3in]{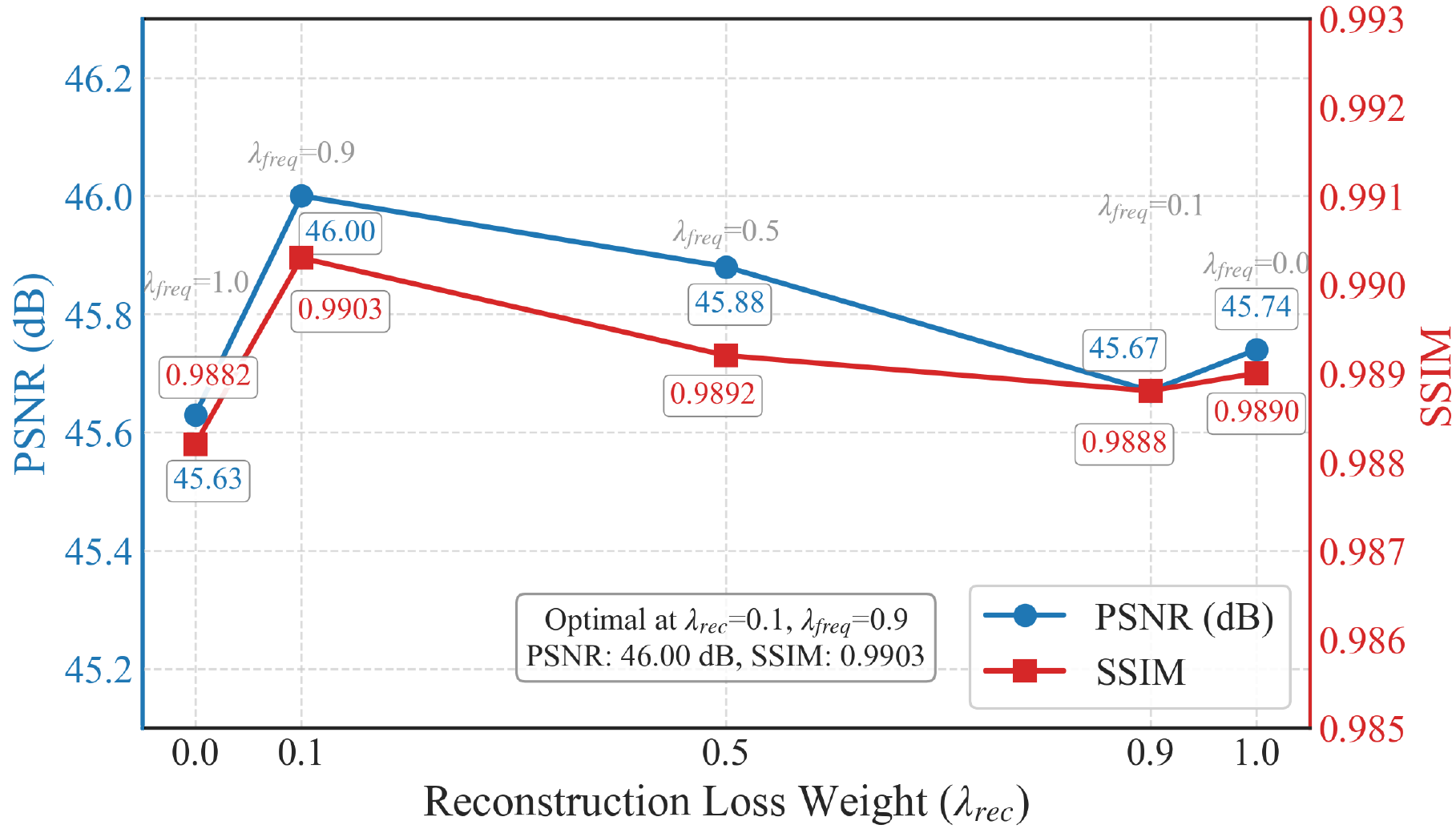}
  \caption{The influence of loss functions under different coefficients while training.} \label{loss_performance}
\end{figure}

First, optimization of loss function coefficients is conducted: grid search determines the optimal weighting for reconstruction loss $\mathcal{L}_{rec}$ and frequency reconstruction loss $\mathcal{L}_{freq}$ ($\mathcal{L}_{Total} = \lambda_{rec} \mathcal{L}_{rec} + \lambda_{freq} \mathcal{L}_{freq}$). The  experimental results are shown in Fig. \ref{loss_performance}. As can be observed that the best SR performance is obtained at $\lambda_{rec} = 0.1$.

\begin{table}[!h]
\centering
\caption{The influence of different channel numbers on the proposed WMSR on the GHRSST dataset.}
\begin{tabular}{ccccc}
\hline\toprule
Number of Channels & PSNR $\uparrow$ & SSIM$\uparrow$ & FLOPs & Params \\ 
\midrule
32   & 44.79 & 0.9753 & 3.799G & 173.051k \\ 
48   & 45.52 & 0.9858 & 10.720G & 376.299k \\ 
64   & 46.04 & 0.9903 & 23.689G & 657.302K \\ 
96   & 46.09 & 0.9916 & 78.582G & 1453K \\ 
128  & 46.22 & 0.9934 & 194.638G & 2560K \\ 
\bottomrule\hline
\end{tabular}
\label{exp_channels}
\end{table}

Subsequently, the impact of channel numbers in WAM modules is investigated. Ablation studies across 32-128 channels (Table \ref{exp_channels}) reveal that the 64-channel configuration achieves optimal balance between PSNR and computational efficiency, reducing parameters by 74\% compared to the 128-channel model while maintaining performance degradation $<$0.2 dB.

\begin{table}[!h]
\centering
\caption{The influence of different number of WAM blocks on the proposed WMSR on the GHRSST dataset}
\begin{tabular}{ccccccc}
\hline\toprule
RWAMG & WAM &  PSNR $\uparrow$  &  SSIM $\uparrow$  &  FLOPs  &  Params  \\ 
\midrule
\multirow{2}{*}{2} & 2 & 40.13 & 0.9801 & 6.191G & 177.371K \\ 
\multirow{2}{*}{}  & 4 & 42.86 & 0.9897 & 12.024G & 337.321K \\ 
\midrule
\multirow{2}{*}{4} & 2 & 42.66 & 0.9894 & 12.180G & 329.724K \\ 
\multirow{2}{*}{}    & 4 & 46.04 & 0.9903 & 23.689G & 657.371K \\ 
\midrule
\multirow{2}{*}{6} & 2 & 45.50 & 0.9906 & 17.856G & 497.343K \\ 
\multirow{2}{*}{}    & 4 & 46.87 & 0.9913 & 35.354G & 977.261K \\ 
\bottomrule\hline
\end{tabular}
\label{exp_numbers}
\end{table}

Finally, module quantity combinations are optimized in Table \ref{exp_numbers}). Comparative experiments with varying RWAMG groups $N_g$ and WAM modules per group $N_m$ confirm that the configuration $[N_g, N_m] = [6, 4]$ yields peak PSNR on the GHRSST dataset. Based on this analysis, the final architecture adopts 24 WAM modules ($4_{\text{groups}} \times 6_{\text{WAM/group}}$) as the baseline.

\begin{table*}[!ht]
\centering
\caption{Super-resolution performance of different methods on the HYCOM Dataset. PSNR and SSIM are calculated in the test dataset. The red font shows the best performance, and the blue font shows the second-best indicators.} \label{table_hycom}
% \resizebox{0.9\linewidth}{!}{
\begin{tabular}{cccccccccccc}
\hline\toprule
\multirow{2}{*}{Method} &\multirow{2}{*}{SR scale} & \multirow{2}{*}{FLOPs} & \multirow{2}{*}{Params} &\multicolumn{2}{c}{NP} &\multicolumn{2}{c}{AGM} &\multicolumn{2}{c}{EWP} &\multicolumn{2}{c}{NIO}\\ 
\cmidrule{5-12}
\multirow{2}{*}{} &\multirow{2}{*}{} & & & PSNR $\uparrow$& SSIM $\uparrow$& PSNR$\uparrow$ & SSIM $\uparrow$& PSNR$\uparrow$ & SSIM $\uparrow$& PSNR $\uparrow$& SSIM $\uparrow$ \\
\midrule
\multirow{3}{*}{EDSR \cite{edsr2017cvpr}} & ×2  &113.850G &  1370K& 36.95 & 0.9837 & 45.63 & 0.9893 & 43.93 & 0.9928 & 37.64 & 0.9813  \\
\multirow{3}{*}{}     &×3   &50.600G &  1370K & 31.34 & 0.9651 & 39.72 & 0.9812 & 37.25 & 0.9736 & 33.89 & 0.9685  \\ 
\multirow{3}{*}{}     &×4   &28.462G &  1370K & 29.77 & 0.9341 & 38.57 & 0.9771 & 32.16 & 0.9332 & 32.31 & 0.9465  \\ 
\cmidrule{2-12}
\multirow{3}{*}{CRAN \cite{cran2021iccv}}&×2   &658.642G &  8001K & 37.63 & 0.9871 & 45.96 & 0.9903 & 44.43 & 0.9950 & 37.92 & 0.9853  \\
\multirow{3}{*}{}    &×3   &292.730G &  8001K & 31.34 & 0.9651 & 39.72 & 0.9812 & 37.25 & 0.9736 & 33.89 & 0.9685  \\ 
\multirow{3}{*}{}    &×4   & 164.660G &  8001K  & 30.42 & 0.9362 & 39.16 & 0.9786 & 36.80 & 0.9776 & 33.02 & 0.9389   \\ 
\cmidrule{2-12}
\multirow{3}{*}{ELAN \cite{elan}}&×2   &118.285G &  1432K& \color{blue}43.54 & \color{red}0.9915 & \color{blue}49.62 & \color{blue}0.9930 & \color{blue}49.03 & \color{blue}0.9946 & 44.07 & 0.9869  \\
\multirow{3}{*}{}    &×3   &52.571G &  1432K & 37.70 & 0.9779 & 47.32 & \color{blue}0.9893 & 45.86 & 0.9920 & 39.79 & 0.9764   \\ 
\multirow{3}{*}{}    &×4   & 29.571G &  1432K &\color{blue}35.73 & \color{blue}0.9653 & \color{blue}45.58 & 0.9852 & \color{blue}44.10 & \color{blue}0.9894 & 37.96 & 0.9650  \\ 
\cmidrule{2-12}
\multirow{3}{*}{DAT \cite{dat2023iccv}} &×2   &421.355G &  5224K& 38.19 & 0.9876 & 47.43 & 0.9920 & 44.91 & 0.9918 & 40.81 & 0.9893  \\
\multirow{3}{*}{}    &×3   & 187.270G &  5224K & 34.81 & 0.9709 & 44.89 & 0.9879 & 39.80 & 0.9837 & 38.23 & 0.9763  \\ 
\multirow{3}{*}{}    &×4   & 105.340G &  5224K & 32.54 & 0.9483 & 42.94 & 0.9823 & 41.31 & 0.9852 & 36.21 & 0.9562  \\ 
\cmidrule{2-12}
\multirow{3}{*}{SFAM \cite{safmn2023iccv}} & ×2   & 458.550G &  5551K & \color{red}43.67 & 0.9905 & 48.31 & 0.9918 & 47.71 & 0.9931 & \color{blue}44.81 & 0.9894  \\
\multirow{3}{*}{}    &×3   & 203.800G &  5551K & \color{blue}39.08 & \color{blue}0.9810 & 47.28 & 0.9894 & \color{red}47.15 & 0.9921 & 40.31 & \color{blue}0.9799  \\ 
\multirow{3}{*}{}    &×4   & 114.637G &  5551K  & 35.13 & 0.9627 & 43.61 & 0.9837 & 43.38 & 0.9885 & 37.09 & 0.9608  \\ 
\cmidrule{2-12}
\multirow{3}{*}{ManbaIR \cite{mambair2024eccv}} &×2   & 169.921G &  2043K & 42.01 & \color{red}0.9915 & 49.42 & 0.9929 & 48.53 & 0.9944 & 43.75 & \color{blue}0.9902  \\% 
\multirow{3}{*}{}        &×3   & 82.634G &  2227K & 37.72 & \color{blue}0.9810 & \color{red}47.57 & \color{red}0.9899 & 46.36 & \color{blue}0.9924 & \color{red}41.52 & 0.9713  \\ 
\multirow{3}{*}{}        &×4   & 55.141G &  2190K & 35.35 & \color{red}0.9673 & 44.86 & \color{blue}0.9853 & 42.75 & 0.9866 & \color{blue}37.99 & \color{red}0.9684  \\ 
\cmidrule{2-12}
\multirow{3}{*}{Proposed WMSR} &×2   & 88.242G &  977K & 43.36 & \color{blue}0.9911 & \color{red}49.91 & \color{red}0.9931 & \color{red}49.65 & \color{red}0.9947 & \color{red}44.93 & \color{red}0.9916  \\
\multirow{3}{*}{}    &×3   & 50.255G &  977K & \color{red}39.83 & \color{red}0.9881 & \color{blue}47.42 & \color{blue}0.9893 & \color{blue}47.13 & \color{red}0.9927 & \color{blue}41.37 & \color{red}0.9874  \\ 
\multirow{3}{*}{}    &×4   & 35.354G &  977K & \color{red}36.20 & \color{red}0.9673 & \color{red}45.87 & \color{red}0.9853 & \color{red}44.82 & \color{red}0.9900 & \color{red}38.26 & \color{blue}0.9662  \\ 
\bottomrule\hline
\end{tabular}
\end{table*}

\subsection{Compare with the state of the art methods}

\begin{table*}[!h]
\footnotesize
\centering
\caption{Super-resolution performance of different methods on the OISST Dataset. PSNR and SSIM are calculated in the test dataset. The red font shows the best performance, and the blue font shows the second-best values.} \label{oisst}
\begin{tabular}{ccccccccc}
\hline\toprule
\multirow{1}{*}{Scale} &{}& EDSR \cite{edsr2017cvpr} & CRAN \cite{cran2021iccv} & ELAN \cite{elan} & DAT \cite{dat2023iccv} & SFAM \cite{safmn2023iccv} & MambaIR \cite{mambair2024eccv} & Proposed WMSR \\ 
\midrule
\multirow{2}{*}{$\times$2}  &PSNR$\uparrow$&	 41.83&42.14&\color{blue}46.88&42.74&45.18&46.31&\color{red}47.07 \\ 
\multirow{2}{*}{}    &SSIM$\uparrow$&	0.9887&0.9905&\color{red}0.9956&0.9913&\color{blue}0.9933&0.9943&\color{red}0.9956 \\ \cmidrule{3-9}
\multirow{2}{*}{$\times$3}  &PSNR$\uparrow$&	36.36&37.42&38.80&37.72&\color{red}39.59&39.02&\color{blue}39.37 \\ 
\multirow{2}{*}{}    &SSIM$\uparrow$&	0.9757&0.9793&0.9869&0.9806&\color{blue}0.9865 & 0.9862&\color{red}0.9889 \\ \cmidrule{3-9}
\multirow{2}{*}{$\times$4} &PSNR$\uparrow$&34.84 & 35.11 & 35.28 & 33.15 & 35.66 & \color{blue}35.87 & \color{red}35.98 \\
\multirow{2}{*}{}    &SSIM$\uparrow$&0.9566 & 0.9578 & 0.9679 & 0.9568 & 0.9669 &\color{blue} 0.9705 & \color{red}0.9712 \\
\bottomrule\hline
\end{tabular}
\end{table*}

To evaluate the super-resolution performance of the proposed WMSR, we compared its performance with six state-of-the-art methods, including EDSR \cite{edsr2017cvpr}, CRAN \cite{cran2021iccv},  ELAN \cite{elan}, DAT \cite{dat2023iccv}, SAFM \cite{safmn2023iccv} and MambaIR \cite{mambair2024eccv}. 

\begin{figure*} [!t]
  \centering
  \includegraphics[width=6in]{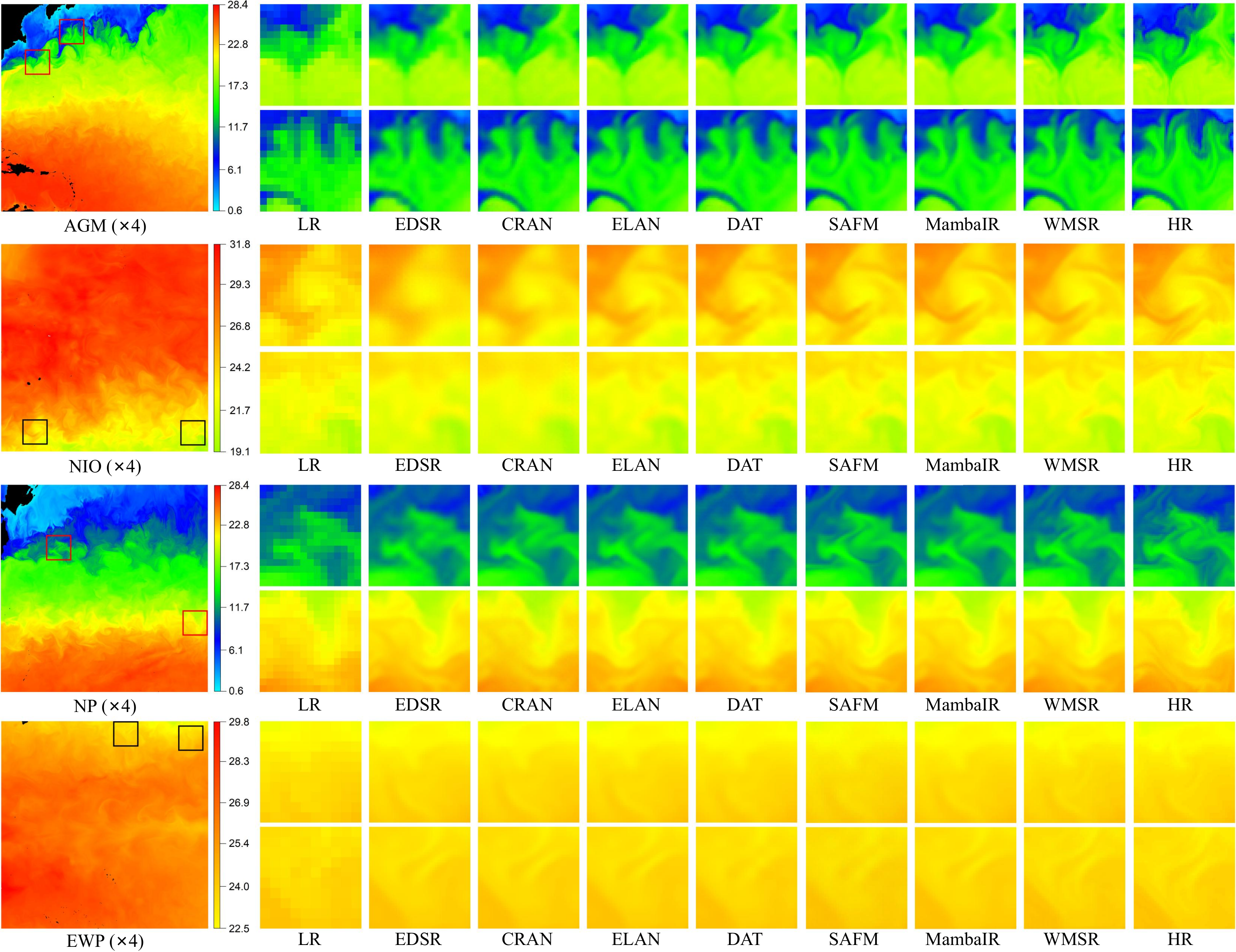}
  \caption{Qualitative analysis of different methods on the HYCOM dataset with super-resolution scale of 4. Each region intercepts fixed data and zooms it to the same size to best observe detail. The four regions are AGM, NIO, NP, and EWP.} \label{fig_hycom_res}
\end{figure*}

\textbf{Results on the HYCOM dataset.}
As shown in Table \ref{table_hycom}, the super-resolution results achieved by different methods across four areas are summarized. Notably, most existing super-resolution models achieve high PSNR and SSIM. The proposed WMSR demonstrates superior performance in 2$\times$ super-resolution tasks, achieving the highest metrics in three of the four regions. For 3$\times$ super-resolution tasks, WMSR maintains optimal performance across all regions. In 4$\times$ super-resolution tasks, WMSR achieves the highest PSNR scores across all oceanic regions. Collectively,  the proposed WMSR delivers the best super-resolution performance.

Considering the computational cost and parameters of each method, compared to the high efficiency but low performance of EDSR and the high performance yet high consumption of SFAM/CRAN, WMSR simultaneously achieves overall optimal performance while maintaining lower resource consumption than EDSR. Compared to ELAN and ManbaIR, which also pursue a balance between computational cost and performance, WMSR delivers superior results with significantly reduced parameters (approximately 30-55$\%$ fewer) and substantially lower computational requirements (approximately 25-50$\%$ less). WMSR not only features lower computational complexity but also leads to improved reconstruction quality on most metrics. It strikes the best balance between performance and efficiency, significantly outperforming other methods.

\begin{figure*}[!ht]
  \centering
  \includegraphics[width=6in]{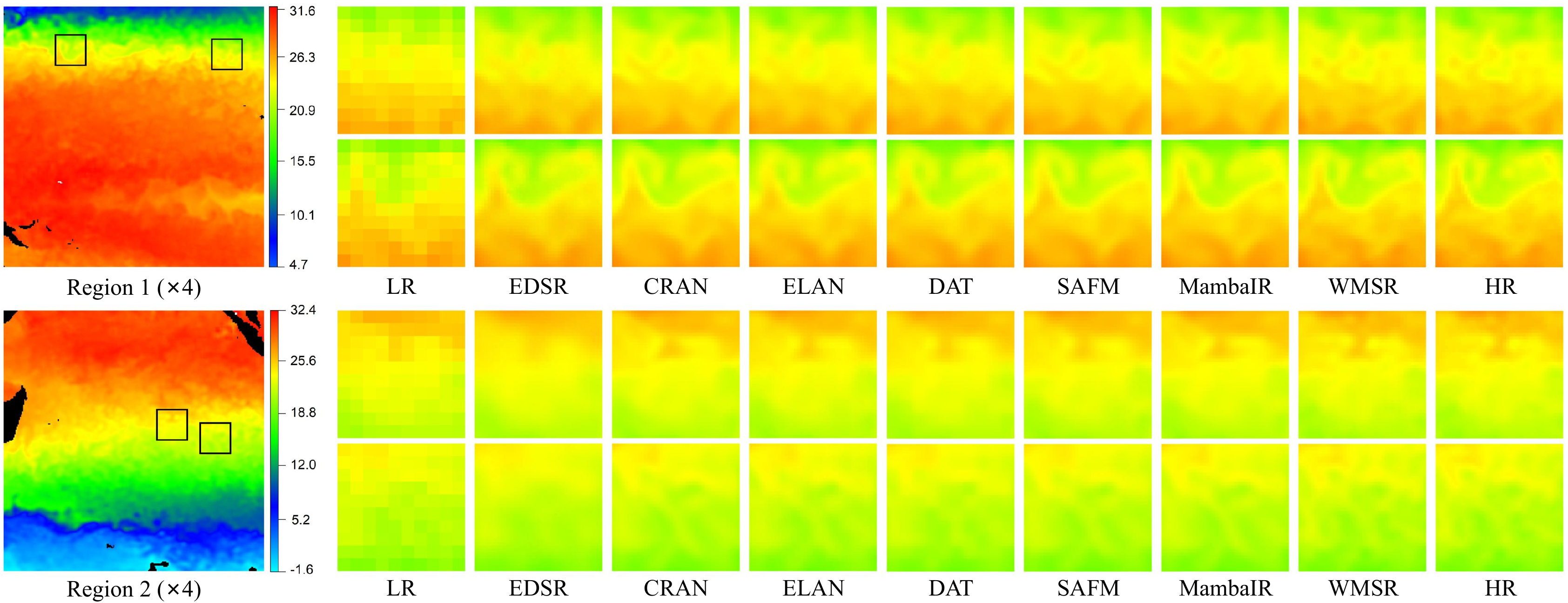}
  \caption{Qualitative analysis of different models on the OISST dataset with super-resolution scale of 4, randomly selected two regions in the test set, each region intercepted a fixed size of data, and scaled to the same size, in order to best observe the details.} \label{fig_oisst_res}
\end{figure*}

Fig. \ref{fig_hycom_res} shows the visualized experimental results of different methods on the HYCOM dataset. Notably, in the first two rows of Fig. \ref{fig_hycom_res}, the proposed WMSR proficiently captures small-scale vortices, providing an ideal spatial reconstruction with minimal smoothing or distortion. Simultaneously, concerning the comparatively smooth low-resolution data presented in the third and fourth rows of Fig. \ref{fig_hycom_res}, all methods can reconstruct certain SST texture details, and the WMSR reconstructs more detailed information.

\begin{table*}[!h]
\footnotesize
\centering
\caption{Super-resolution performance of different methods on the GHRSST Dataset. PSNR and SSIM are calculated in the test dataset. The red font shows the best performance, and the blue font shows the second-best indicators.} \label{ghrsst}
\begin{tabular}{ccccccccc}%\hline
\hline\toprule
\multirow{1}{*}{Scale} & & EDSR \cite{edsr2017cvpr} & CRAN \cite{cran2021iccv} & ELAN \cite{elan} & DAT \cite{dat2023iccv} & SFAM \cite{safmn2023iccv} & MambaIR \cite{mambair2024eccv} & Proposed WMSR \\ 
\cmidrule{1-9}
\multirow{2}{*}{$\times$2}  &PSNR$\uparrow$& 49.36 & 48.37 & 51.57 & 51.13 & 50.20 & \color{blue}52.42 & \color{red}53.67 \\ 
\multirow{2}{*}{}    &SSIM$\uparrow$& 0.9911 & 0.9960 & 0.9971 & \color{blue}0.9973 & 0.9942 & \color{blue}0.9973 & \color{red}0.9978 \\ 
\cmidrule{3-9}
\multirow{2}{*}{$\times$3}  &PSNR$\uparrow$& 45.46 & 46.22 & \color{blue}49.87 & 48.97 & 48.42 & 48.56 & \color{red}50.36\\ 
\multirow{2}{*}{}    &SSIM$\uparrow$& 0.9941 & 0.9926 & \color{blue}0.9958 & \color{blue}0.9958 & 0.9934 & 0.9955 & \color{red}0.9960 \\ 
\cmidrule{3-9}
\multirow{2}{*}{$\times$4}  &PSNR$\uparrow$& 42.99 & 44.21 & 46.79 & 46.06 & 45.11 & \color{blue}47.46 & \color{red}47.89\\
\multirow{2}{*}{}    &SSIM$\uparrow$&0.9721 & 0.9901 & 0.9921 & 0.9909 & 0.9875 & \color{red}0.9931 & \color{blue}0.9928    \\ 
\bottomrule\hline
\end{tabular}
\end{table*}

\begin{figure*}[!ht]
  \centering
  \includegraphics[width=6in]{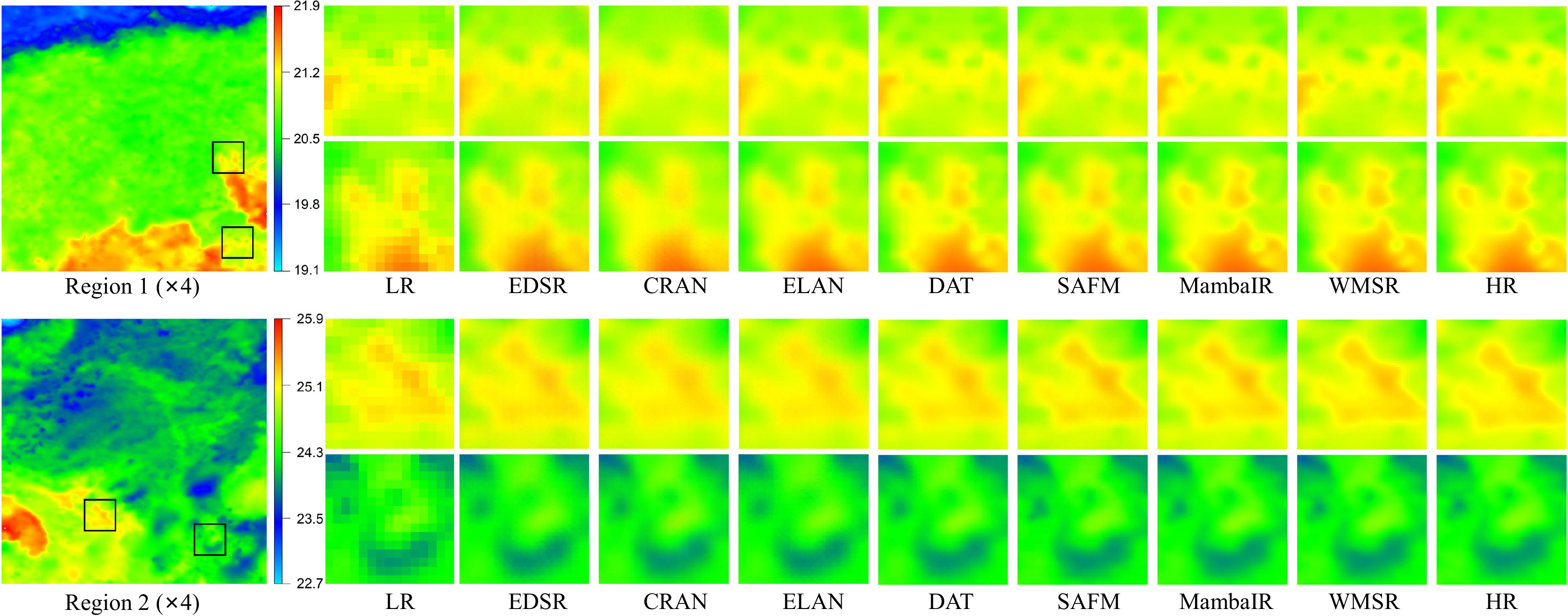}
  \caption{Qualitative analysis of different models on the GHRSST dataset, super-resolution scale of 4, randomly selected two regions in the test set, each region intercepted a fixed size of data, and scaled to the same size, in order to best observe the details.} \label{ghr}
\end{figure*}

\textbf{Comparison of super-resolution results on OISST dataset.} Table \ref{oisst} presents the super-resolution results of various methods evaluated on the OISST dataset. In the $2 \times$ super-resolution experiment, although most super-resolution models achieved high PSNR and SSIM metrics, WMSR outperforms the other methods by a considerable margin. In the $3 \times$ super-resolution experiment, WMSR delivered the highest SSIM value. For the $4 \times$ super-resolution task, the proposed WMSR secured the top spot in PSNR and SSIM metrics, highlighting a significant advantage, while MambaIR, with its state-space modeling, secured suboptimal performance. 

Fig. \ref{fig_oisst_res} presents the comparative analysis of various super-resolution methods.  For the intricate ocean current area (Fig. \ref{fig_oisst_res} Region 1), the reconstruction produced by WMSR exhibits minimal artifacts and distortions, most accurately reproducing the texture details of the SST.  In Fig. \ref{fig_oisst_res} Region 2, other super-resolution methods clearly exhibit over-smoothing, resulting in low fidelity to the HR reference. Conversely, WMSR generates images that preserve consistent texture details, demonstrating the superior capability of the proposed method in detail reconstruction.

\textbf{Comparison of super-resolution results on GHRSST dataset.}
Table \ref{ghrsst} presents super-resolution results on the GHRSST dataset across comparative methods. The proposed WMSR demonstrates superior overall performance, achieving state-of-the-art PSNR metrics in both 2$\times$ and 3$\times$ tasks. For 4$\times$ reconstruction, WMSR attains optimal PSNR values while maintaining competitive SSIM performance. These results establish WMSR as an effective solution for sea surface temperature super-resolution applications.%The results indicate that the proposed WMSR demonstrates superior super-resolution performance across most metrics.

Fig.\ref{ghr} compares super-resolution outputs generated by competing methodologies. Visual analysis reveals that conventional approaches exhibit boundary ambiguity during detail reconstruction(Fig. \ref{ghr} region 1), whereas the proposed WMSR model demonstrates superior structural fidelity to high-resolution references. Crucially, as evidenced in Fig. \ref{ghr} region 2, WMSR achieves precise boundary delineation without introducing spurious artifacts, which is common failure observed in comparative methods. %These visual representations effectively highlight the advantages of the proposed WMSR.

\begin{table*}[!h]
\centering
\caption{The ablation experiments on the HYCOM Dataset. PSNR and SSIM values are calculated in the test dataset.} \label{ablu}
\resizebox{0.95\textwidth}{!}{
\begin{tabular}{cccccccccccccc}
\hline\toprule
\multirow{2}{*}{} & \multirow{2}{*}{HFEM} & \multirow{2}{*}{LFSSM} & \multirow{2}{*}{DWT}& \multirow{2}{*}{Flops} & \multirow{2}{*}{param}  & \multicolumn{2}{c}{NP} & \multicolumn{2}{c}{WA} & \multicolumn{2}{c}{EWP} & \multicolumn{2}{c}{NIO} \\ 
\cmidrule{7-14}
 & & & & & & PSNR$\uparrow$ & SSIM $\uparrow$& PSNR$\uparrow$ & SSIM$\uparrow$ & PSNR$\uparrow$ & SSIM$\uparrow$ & PSNR$\uparrow$ & SSIM$\uparrow$ \\ 
\midrule
Case1  & \checkmark & \checkmark & \checkmark &{35.354G} & 977.371K&  34.95 & 0.9597 & 44.80 & 0.9841  & 43.63 & 0.9889 & 37.84 & 0.9622 \\
Case2  &  & \checkmark & \checkmark  &{34.213G} &  {866.348K}&  34.68 & 0.9586 & 44.52 & 0.9834  & 43.21 & 0.9872 & 37.07 & 0.9593 \\
Case3  & \checkmark &  & \checkmark  &{32.671G }& {462.043K} &  33.42 & 0.9526 & 43.93 & 0.9821  & 42.93 & 0.9868 & 36.87 & 0.9612 \\
Case4  & \checkmark & \checkmark &   &{33.443G} & {498.715K}&  34.75 & 0.9588 & 44.40 & 0.9827  & 43.54 & 0.9868 & 37.14 & 0.9590 \\
\cmidrule{1-14}
Case5  &  &  & \checkmark  &{31.530G }&{351.020K}&  32.38 & 0.9410 & 42.26 & 0.9802  & 41.45 & 0.9847 & 35.61 & 0.9522 \\
Case6  &  & \checkmark &  &{33.443G} &{498.715K}&  34.27 & 0.9573 & 44.31 & 0.9798  & 43.21 & 0.9843 & 37.24 & 0.9602 \\
Case7  & \checkmark &  &  &{33.237G}&{451.968K}&  33.37 & 0.9432 & 43.62 & 0.9699  & 42.83 & 0.9782 & 36.43 & 0.9415 \\
\cmidrule{1-14}
Case8  & w/o PDC & \checkmark & \checkmark &{34.993G} & {942.961K}&  34.72 & 0.9583 & 44.69 & 0.9841  & 42.97 & 0.9875 & 37.30 & 0.9621 \\
Case9  & \checkmark & w/o 2D SSM & \checkmark & {33.722G} & {664.027K}&   34.50 & 0.9579 & 44.62 & 0.9842  & 42.91 & 0.9870 & 37.24 & 0.9613 \\
\bottomrule\hline
\end{tabular}}
\end{table*}

\subsection{Ablation Study}

To determine the contributions of various components in the WMSR, including HFEM, LFSSM, and DWT, we carried out a series of ablation experiments, with the specific experimental results outlined in Table \ref{ablu}. For efficiency, all models were trained for 20 epochs, with $4 \times$ super-resolution task. 

Ablation experiments confirm the critical role of low-frequency information in super-resolution performance. Case 3 demonstrates that removing the LFSSM module causes significant PSNR degradation (-1.53 dB in NP). Cases 1, 3, and 9 collectively prove that the 2D SSM module is essential for high performance despite its substantial parameter cost. Cases 1 and 2 reveal that the HFEM enhances detail reconstruction through high-frequency prior embedding while improving image similarity metrics. Case 8 verifies that the PDC requires minimal parameters, highlighting the inherent efficiency of re-parameterization. Crucially, simultaneous removal of both frequency components (Cases 5–7) induces the most severe performance drop, underscoring the necessity of dual-path synergy. Comparison of Cases 1 and 4, as well as Cases 2 and 6, reveals that the removal of the DWT operation significantly reduces model parameters. However, the computational load does not exhibit a commensurate reduction. By comparing Case 8 and Case 1, the PDC component contributed an average of 0.39dB under extremely low parameters, demonstrating the crucial role of gradient enhancement in detail retention,  highlighting the inherent efficiency of re-parameterization.   Combined with Cases 1 and 9, we can find 2D-SSM contributed an average performance gain of approximately 0.49dB respectively, indicating the effectiveness of 2D-SSM in explicitly capturing long-range dependencies through structured state Spaces.

\section{Conclusion and Future Work}

In this paper, we designed the WMSR framework for super-resolution of satellite-derived SST data. The framework aims to address the SST restoration problem by leveraging a selective scan mechanism. Initially, input features are decomposed into low and high-frequency components using the discrete wavelet transform. To capture global information within the input data, we introduce the Low-Frequency Selective Scan Module, which exploits the global modeling capabilities of 2D-SSM to preserve critical thermal information contained in the low-frequency component. Concurrently, to enhance spatial details, we design the HFEM, which employs PDC for refining high-frequency features, thereby achieving accurate and sharp texture reconstruction. Extensive experiments conducted on three benchmark datasets validated the effectiveness of the proposed WMSR framework. The results further demonstrate that WMSR successfully reconstructs complex structures associated with ocean phenomena from degraded satellite-derived SST data.

In the future, we plan to explore advanced wavelet bases to better discriminate low/high-frequency features in satellite SST data, enhancing reconstruction fidelity. This may further improve the performance of SST data super-resolution. Additionally, we aim to extend the application scope of the WMSR framework to other satellite-derived data, such as sea surface height and ocean color data. By adapting the selective scan mechanism, we can potentially make more comprehensive understanding of ocean dynamics.

\bibliography{source}
\bibliographystyle{IEEEtran}
\end{document}